\documentclass
[final,1p,times]{elsarticle}
\journal{E}
\usepackage{tikz-cd}
\usepackage{extpfeil}

\usepackage{amssymb}
\usepackage{amsthm}
\usepackage{latexsym}
\usepackage{amsmath}
\usepackage{color}
\usepackage{graphicx}
\usepackage{indentfirst}
\usepackage{mathrsfs}
\usepackage{lipsum}
\usepackage{comment}
\usepackage{physics}
\usepackage{subcaption}

\usepackage[pdfencoding=auto,psdextra,hidelinks]{hyperref}
\usepackage{bookmark}

\usepackage{todonotes}
\usepackage{xcolor}

\allowdisplaybreaks

\renewcommand{\thefootnote}{\arabic{footnote}}
\newtheorem{theorem}{\color{black}\indent \textbf{Theorem}}[section]

\newtheorem{proposition}{\color{black}\indent Proposition}[section]
\newtheorem{definition}{\color{black}\indent Definition}[section]
\newtheorem{remark}{\color{black}\indent Remark}[section]

\newtheorem{example}{\color{black}\indent Example}[section]

\newcommand\blfootnote[1]{% 
	\begingroup 
	\renewcommand\thefootnote{}\footnote{#1}% 
	\addtocounter{footnote}{-1}% 
	\endgroup 
}

\begin{document}
	\begin{frontmatter}

		\title{Integrability of Hamiltonian systems on COLCS Manifold}
\author{{ \blfootnote{$^{*}$Corresponding author at: School of Mathematics, Jilin University, Changchun 130012, People’s Republic of China} Antonio J. Pan-Collantes$^{a}$\footnote{ E-mail address : antonio.pan@uca.es},
				~Xuefeng Zhao$^{b,*}$ \footnote{E-mail address :  zhaoxuef@jlu.edu.cn}}  \\
			{$^{a}$Department of Mathematics, Universidad de C\'{a}diz, Puerto Real, Spain\\
			{$^{b}$College of Mathematics, Jilin University,} {Changchun 130012, P. R. China}.}
		}

% 		\begin{abstract}
% This paper develops an integrability framework for Hamiltonian dynamics on a class of odd-dimensional manifolds known as con-locally conformal symplectic (COLCS) manifolds. A COLCS manifold is a quadruple $(M,\Omega,\theta,\eta)$, where $\Omega$ is a $2$-form, $\theta$ and $\eta$ are closed $1$-forms satisfying $d\Omega = \theta\wedge\Omega$, and $\Omega$ is nondegenerate on the codimension-one distribution $\ker\eta$. The COLCS bracket on smooth functions is introduced and its algebraic properties are analyzed. Although the bracket does not define a Poisson structure on the whole algebra $C^\infty(M)$, it induces Poisson structures on certain natural subalgebras, such as the $\theta$-strong and $R$-strong functions. A Lie-type integrability theorem is proved for $\theta$-strong Hamiltonian vector fields: under suitable first-integral and commutation conditions, the dynamics can be integrated by quadratures on a common level set. The notion of a scaling symmetry of degree $(\Lambda,\beta,\gamma)$ is introduced and studied, which provides a more general class of symmetries. This work offers new geometric tools for the study of time-dependent, dissipative, or twisted Hamiltonian systems, thereby extending the classical integrability theory of symplectic manifolds.
% 		\end{abstract}

\begin{abstract}
We develop an integrability framework for Hamiltonian dynamics on
\emph{con-locally conformal symplectic} (COLCS) manifolds, odd-dimensional
quadruples $(M,\Omega,\theta,\eta)$ where $d\Omega=\theta\wedge\Omega$,
a closed $1$-form $\eta$ determines a codimension-one distribution on which
$\Omega$ is non-degenerate, and $R$ is the Reeb vector field satisfying
$\iota_R\Omega=\iota_R\theta=0$, $\iota_R\eta=1$.
This class simultaneously generalises LCS and cosymplectic manifolds and
provides a natural arena for time-dependent Hamiltonian systems with
twisted differential $d^\theta=d-\theta\wedge$.

The COLCS bracket is introduced on $C^\infty(M)$ and shown to be a Lie bracket
that induces Poisson structures on the subalgebras of $\theta$-strong
($\theta(X_H)=0$) and $R$-strong ($R(H)=0$) functions.
A Lie-type integrability theorem is then established: given $2n-k$ functionally
independent first integrals with $k$ of them $\theta$-strong and generating a
solvable Lie algebra under the COLCS bracket, the flow is integrable by
quadratures on the common level set.
Finally, scaling symmetries of degree $(\Lambda,\beta,\gamma)$, defined by
$L_X\Omega=\beta\Omega$, $L_XH=\Lambda H$, $L_X\eta=\gamma\eta$,
are studied: they rescale $X_H$ by $(\Lambda-\beta)$, generate families of
first integrals, and imply structural primitives for $\Omega$ and $\eta$.
\end{abstract}

	\end{frontmatter}
	\section{Introduction}
		Hamiltonian mechanics on symplectic manifolds provides a geometric language for conservative dynamics, and its integrability theory (Liouville--Arnold , Lie, Magri non-Hamiltonian, $C^\infty$ integrability) is by now classical \cite{Arnold3,Azuaje1,Fuchssteiner,Liouville,Magri,Olver,pancinf-sym,Zhao}. In many contexts, however, the relevant phase space carries a structure that is not globally symplectic: locally conformal symplectic (LCS) manifolds \cite{Azuaje,Gray,Lee2,Lefebvre,Libermann,Zhao2} appear naturally in the study of conformal and dissipative systems, and they also arise from reduction and from the geometry of twisted differentials \cite{Carinena,Morando,Zhao3}. In parallel, time-dependent Hamiltonian systems are conveniently encoded on manifolds of one dimension higher by adjoining a time variable.

		The goal of this paper is to develop an integrability framework for Hamiltonian dynamics on a class of odd-dimensional manifolds that combines these two features. We work with 
		\emph{con-locally conformal symplectic} (COLCS) manifolds, i.e.
		quadruples $(M,\Omega,\theta,\eta)$ where $\Omega$ is a $2$-form satisfying $d\Omega=\theta\wedge\Omega$ for a closed Lee form $\theta$, and where a closed $1$-form $\eta$ determines a codimension-one distribution $\xi=\ker\eta$ on which $\Omega$ is non-degenerate, together with a distinguished Reeb vector field $R$ characterized by $\iota_R\Omega=0$, $\iota_R\theta=0$ and $\iota_R\eta=1$. This setup contains, as basic examples, products $S^1\times N$ where $(N,\omega,\theta)$ is LCS, and it provides a natural geometric arena for time-dependent Hamiltonian systems with a twisted differential $d^\theta=d-\theta\wedge$.

		After introducing Hamiltonian and evolution vector fields associated to a Hamiltonian function $H\in C^\infty(M)$, we define a COLCS bracket on functions and analyze its algebraic properties. In contrast to the symplectic case, the bracket is not Poisson on the full algebra $C^\infty(M)$ in general. We identify natural subalgebras (the $\theta$-strong and strongest function classes) on which the bracket does induce a Poisson structure, and we clarify the role played by the Reeb dynamics in the obstruction to the Jacobi identity.

		We emphasize that the time-dependent dynamics is encoded by the evolution vector field $E_H=X_H+R$, while our integrability results in this draft are formulated for the Hamiltonian vector field $X_H$ (and its restrictions to suitable invariant submanifolds). Since first-integral and symmetry conditions for $E_H$ differ from those for $X_H$ in general, we will always state explicitly when a claim concerns $E_H$.

		\sloppy
		The central results concern integrability. Inspired by Lie's theorem and by the non-Hamiltonian integrability scheme of Bogoyavlenskij, we establish a Lie-type integrability theorem for $\theta$-strong Hamiltonian vector fields on COLCS manifolds: under suitable commutation and first-integral hypotheses expressed in terms of the COLCS bracket, the dynamics can be integrated by quadratures on an appropriate common level set. Finally, we study scaling symmetries for time-dependent Hamiltonian systems in the COLCS setting, extending the notion introduced for time-dependent Hamiltonian systems, and we relate these symmetries to first integrals and structural properties of the twisted differential.

		The paper is organized as follows. Section~\ref{colcs-manifolds} introduces COLCS manifolds, establishes the Darboux-type normal form, and defines the associated Hamiltonian and evolution vector fields. The COLCS bracket is then defined on $C^\infty(M)$ and its algebraic properties are analyzed: it is shown to be a Lie bracket on the full algebra, and to induce a Poisson structure on the natural subalgebras of $\theta$-strong, $R$-strong, and strongest functions. Section~\ref{lie-integrability} develops the integrability theory. After recalling Lie's theorem and the non-Hamiltonian integrability framework of Bogoyavlenskij, we prove a Lie-type integrability theorem (Theorem~\ref{TZ}) for $\theta$-strong Hamiltonian vector fields: under suitable commutation and first-integral hypotheses expressed via the COLCS bracket, the dynamics can be integrated by quadratures on a common level set. Section~\ref{scaling-symmetries} introduces and studies scaling symmetries of degree $(\Lambda,\beta,\gamma)$ for Hamiltonian systems on COLCS manifolds. We show that such a symmetry rescales the Hamiltonian vector field by $(\Lambda-\beta)$, derive their Lie-algebraic closure properties, and establish that their existence implies primitives for both $\Omega$ under the twisted differential $d^\theta$ and for $\eta$.

\section{COLCS manifolds and associated Hamiltonian vector fields}\label{colcs-manifolds}

In this section we lay the geometric foundations for the remainder of the paper.
We introduce the class of \emph{con-locally conformal symplectic} (COLCS) manifolds, establish their local normal form via a Darboux-type theorem, and define the Hamiltonian and evolution vector fields that encode the dynamics.
The section closes with the COLCS bracket and an analysis of the function subalgebras on which it induces a genuine Poisson structure.

Recall that a \emph{locally conformal symplectic} (LCS) manifold is a pair $(N,\omega)$, where $\omega$ is a non-degenerate $2$-form on an even-dimensional manifold $N$ satisfying $d\omega = \theta\wedge\omega$ for a closed \emph{Lee form} $\theta$ \cite{Chantraine,Esen,Lee2,Libermann,Vaisman}.
COLCS manifolds arise by adjoining, to such a structure, a contact-like direction encoded by a closed $1$-form $\eta$ and a corresponding Reeb vector field; the resulting odd-dimensional geometry simultaneously generalises LCS manifolds and cosymplectic manifolds, and provides a natural arena for time-dependent Hamiltonian systems with a twisted differential.

	\begin{definition}
		Let $M$ be a smooth manifold of dimension $2n+1$. A \textbf{COLCS manifold} (short for \emph{Con-Locally Conformal Symplectic manifold}) is a quadruple $(M, \Omega, \theta, \eta)$ where:
		\begin{itemize}
			\item $\Omega$ is a 2-form on $M$;
			\item $\theta$ and $\eta$ are closed 1-forms on $M$;
			\item The following structure equations are satisfied:
			\[
			d\Omega = \theta \wedge \Omega, \qquad \Omega^n \wedge \eta \neq 0 \quad \text{everywhere on } M;
			\]
			\item The tangent bundle $TM$ admits a splitting
			\[
			TM = \mathbb{R} \cdot R \oplus \xi,
			\]
			where $R$ is called a Reeb vector field which satisfies
			\[
			\iota_R \Omega = 0, \qquad \iota_R \theta = 0, \qquad \iota_R \eta = 1;
			\]
			\item The restriction $\Omega|_{\xi}$ is non-degenerate and the restriction $\eta|_\xi=0.$
		\end{itemize}
	\end{definition}
	\begin{remark}\label{Bao}
	    According to the definition of a COLCS manifold $(M, \Omega, \theta, \eta)$, the Reeb vector field preserves the COLCS structure $(\Omega, \theta, \eta)$. In other words, these three structures are invariant under the flow of the Reeb vector field. In fact,
$$L_R\Omega=di_R\Omega+i_Rd\Omega=i_R(\theta\wedge\Omega)=0,\quad L_R\theta=di_R\theta+i_Rd\theta=0,\quad L_R\eta=di_R\eta+i_Rd\eta=di_R\eta=d1=0.$$
\end{remark}

Indeed, the class of COLCS manifolds is abundant. In the following, we construct explicit examples to illustrate this.
\begin{example}\label{E1}
		Let $N$ be a $2n$-dimensional manifold equipped with a locally conformal symplectic (LCS) structure, i.e., a 2-form $\omega \in \Omega^2(N)$ and a closed 1-form $\theta \in \Omega^1(N)$ satisfying
		\[
		d\omega = \theta \wedge \omega.
		\]
		Let $M = S^1 \times N$, and denote by $t$ the coordinate on $S^1$. Consider the projection $\pi: M \to N$, and define the following forms on $M$:
		\[
		\Omega := \pi^* \omega, \qquad \theta' := \pi^* \theta, \qquad \eta := dt.
		\]
		Then we have
		\[
		d\Omega = \pi^*(d\omega) = \pi^*(\theta \wedge \omega) = \theta' \wedge \Omega,
		\]
		and the volume form
		\[
		\Omega^n \wedge \eta = (\pi^*\omega)^n \wedge dt \neq 0
		\]
		since $\omega$ is non-degenerate on $N$.
		
		Therefore, $(M, \Omega, \theta', \eta) = (S^1 \times N, \pi^* \omega, \pi^* \theta, dt)$ defines a CoLCS structure on $M$.
	\end{example}
\begin{remark}
		For every COLCS manifold $(M,\Omega,\theta,\eta),$ we can see that $(M,\Omega,\theta)$ is a locally conformal presymplectic manifold \cite{Chen}.
	\end{remark}
We now present a Darboux-type theorem for COLCS manifolds, which characterizes the local behavior of COLCS structures.
	\begin{theorem}[Darboux-type Theorem for COLCS Manifolds]\label{Darboux}
		Let $(M, \Omega, \theta, \eta)$ be a COLCS manifold.
		Then for every point $p \in M$, there exists a coordinate chart $(x^1, y^1, \dots, x^n, y^n, t)$ centered at $p$ and a smooth function $f$ such that in these coordinates:
		\[
		\eta = dt, \quad \theta = df, \quad \Omega = e^f \sum_{i=1}^n dx^i \wedge dy^i.
		\]
	\end{theorem}
	
	\begin{proof}
		Since $\theta$ is closed, for any point $p\in M,$ by the Poincaré lemma, there exists a smooth function $f$ defined in a neighborhood of $p$ such that $\theta = df$.
		
		Define a new 2-form by
		\[
		\tilde{\Omega} := e^{-f} \Omega.
		\]
		We compute its exterior derivative:
		\[
		\begin{aligned}
			d\tilde{\Omega} &= d(e^{-f} \Omega) = -e^{-f} df \wedge \Omega + e^{-f} d\Omega \\
			&= -e^{-f} \theta \wedge \Omega + e^{-f} (\theta \wedge \Omega) = 0.
		\end{aligned}
		\]
		Thus, $\tilde{\Omega}$ is a closed 2-form.
		
		Moreover, we have:
		\[
		\tilde{\Omega}^n \wedge \eta = e^{-nf} \Omega^n \wedge \eta \neq 0,
		\]
		so $\tilde{\Omega}$ is non-degenerate on the codimension-one distribution $\xi := \ker \eta$.
		
		This shows that $(\tilde{\Omega}, \eta)$ defines a cosymplectic structure in a neighborhood of $p$. It is known (see \cite{deleon}) that in this setting, there exist local coordinates
		\[
		(x^1, y^1, \dots, x^n, y^n, t)
		\]
		such that
		\[
		\tilde{\Omega} = \sum_{i=1}^n dx^i \wedge dy^i, \quad \eta = dt.
		\]
		
		Recalling that $\Omega = e^f \tilde{\Omega}$ and $\theta = df$, we conclude that in these coordinates:
		\[
		\Omega = e^f \sum_{i=1}^n dx^i \wedge dy^i, \quad \eta = dt, \quad \theta = df,
		\]
		as desired.
	\end{proof}
	\begin{remark}
		In canonical coordinates, we have \( R = \frac{\partial}{\partial t} \).
	\end{remark}

	\begin{definition}
		Let $H \in C^\infty(M)$ be a smooth function. The \textbf{Hamiltonian vector field} $X_H$ associated to $H$ is the unique vector field satisfying
		\[
		\iota_{X_H} \Omega = d^\theta H - R(H)\, \eta,\quad \eta (X_H)=0.
		\]
			The assignment \( H \mapsto X_H \) is linear, that is,
		\[
		X_{H + \alpha F} = X_H + \alpha X_F,\quad \forall H,F\in C^\infty(M),\;\alpha\in\mathbb R.
		\]
		
		The \textbf{Evolution vector field $E_H$} associated to $H$ is the unique vector field satisfying
		\[
		\iota_{E_H} \Omega = d^\theta H - R(H)\, \eta,\quad \eta (E_H)=1.
		\]
		where the \emph{twisted differential} $d^\theta$ is defined by
		\[
		d^\theta := d - \theta \wedge.
		\]
	\end{definition}

	\begin{remark}
		Given $H\in C^\infty(M),$ it is clear that the dynamics of a system on $(M,\Omega,\theta,\eta)$ with Hamiltonian function $H$ is given by the evolution vector field 
		$E_H=X_H+R.$ In coordinates, according to Theorem \ref{Darboux},	the trajectories \( \psi(t) = (x^1(t), \dots, x^n(t), y_1(t), \dots, y_n(t), t) \) of the evolution vector field are the integral curves of \( E_H \), which satisfy the equations of motion
		\[
		\dot{x}^i =e^f \frac{\partial H}{\partial y_i}-e^fH \frac{\partial f}{\partial y_i}, \quad \dot{y}_i = -e^f\frac{\partial H}{\partial x^i}+e^fH\frac{\partial f}{\partial x^i}, \quad \dot{t} = 1,
		\]
		where $f$ is the function such that $df=\theta.$
		
		The evolution of a function \( g \in C^\infty(M) \) (an observable) along the trajectories of the evolution vector field is given by
		\[
		\dot{g} = L_{E_H} g = E_H g = X_H g + R g.
		\]
	\end{remark}
	\begin{remark}\label{R:scope-integrability}
		Throughout the paper, our integrability statements are formulated for the Hamiltonian vector field $X_H$ (characterized by $\eta(X_H)=0$), since this is the object naturally controlled by the COLCS bracket and is tangent to $\ker\eta$ (e.g. to the hypersurfaces $\Sigma_c=\{\eta=c\}$ in local Darboux charts). The evolution vector field $E_H$ governs the full time-dependent dynamics, but its first integrals and symmetry conditions are, in general, different from those of $X_H$.
	\end{remark}
	\begin{definition}
			We say that a function \( f \in C^\infty(M) \) is a first integral of the Hamiltonian vector field $X_H$ if  \( L_{X_H} f = 0 \). $f$  is a first integral of the evolution vector field $E_H$ if  \( L_{E_H} f = 0 \).
	\end{definition}
	\begin{remark}
		We can see that 
		$$0=\iota_{X_H}\iota_{X_H}\Omega=X_H(H)-H\theta(X_H)-R(H)\eta(X_H)=X_H(H)-H\theta(X_H), $$
		hence $X_H(H)=H\theta(X_H)$, which implies $H$ may not be first integral of $X_H.$ Moreover,  
		$$0=\iota_{E_H}\iota_{E_H}\Omega=E_H(H)-H\theta(E_H)-R(H)\eta(E_H), $$
		thus $E_H(H)=H\theta(E_H)+R(H)\eta(E_H),$ which implies $H$ may also not be first integral of $E_H.$ 
	\end{remark}
	\begin{example}
		Let $M=\mathbb R^4\times S^1\ni(x,y,w,z,t),\Omega=e^x(dx\wedge dy+dw\wedge dz),\theta=dx,\eta=dt$ we can see that 
		$$d\Omega=e^x dx\wedge dw\wedge dz=\theta\wedge\Omega,\quad d\theta=0,\quad \Omega^2\wedge \eta=2e^{2x}dx\wedge dy\wedge dw\wedge dz\neq 0,$$
		$$R=\frac{\partial}{\partial t},\quad\iota_R\Omega=0,\quad \iota_R\theta=0,\quad \iota_R\eta=1.$$
		Hence, $(\mathbb R^4,\Omega,\theta)$ is a LCS manifold and $(M,\Omega,\theta,\eta)$ is a COLCS manifold. Let $H=e^x(tx+y+w+z),$ then by 
		$$\iota_{X_H}\Omega=d^\theta H-R(H)\eta,\quad \eta(X_H)=0,$$
		we can get the Hamiltonian vector field $X_H=\frac{\partial}{\partial x}-t\frac{\partial}{\partial y}+\frac{\partial}{\partial w}-\frac{\partial}{\partial z}$,  which is a time-dependent Hamiltonian vector field on COLCS manifold $(M,\Omega,\theta,\eta).$ Moreover, by
		\[
		\iota_{E_H} \Omega = d^\theta H - R(H)\, \eta,\quad \eta (E_H)=1,
		\]
		we can get the Evolution vector field $E_H=\frac{\partial}{\partial t}+\frac{\partial}{\partial x}-t\frac{\partial}{\partial y}+\frac{\partial}{\partial w}-\frac{\partial}{\partial z}.$
	\end{example}
	Now, we define a COLCS bracket on a COLCS manifold.
	
	\begin{definition}
		For any functions \( f, g \in C^\infty(M) \) on COLCS manifold $(M,\Omega,\theta,\eta)$, the COLCS bracket of \( f \) and \( g \) is defined by
		\[
		\{f, g\}_{\text{COLCS}} = \Omega(X_f, X_g),
		\]
		where \( X_f \) and \( X_g \) are the Hamiltonian vector fields corresponding to the functions \( f \) and \( g \), respectively. From now on, for simplicity, we denote \( \{f, g\}_{\text{COLCS}} \) by \( \{f, g\} \).
	\end{definition}
	
	Next, we explore the properties of the COLCS bracket defined above.
	\begin{proposition}\label{PR}
		For any function $f,$ the following equation hold:
		$$[R,E_f]=[R,X_f]=X_{R(f)}.$$
	\end{proposition}
	\begin{proof}
		We can see that 
		$$L_R\Omega=\iota_Rd\Omega+d\iota_R\Omega=\iota_R(\theta\wedge\Omega)=0,$$
		thus
		\begin{align*}
			\iota_{[R,X_f]}\Omega&=\iota_{[R,X_f]}\Omega+\iota_{X_f}L_R\Omega\\
			&= L_{R}(\iota_{X_f}\Omega)\\
			&=L_R(d^\theta f-R(f)\eta)\\
			&=L_R(df-f\theta-R(f)\eta)\\
			&=dR(f)-R(f)\theta-R(R(f))\eta,\\
			\iota_{[R,X_f]}\eta&=L_R(\iota_{X_f}\eta)-\iota_{X_f}L_R\eta=0,
		\end{align*}
		which means that $[R,X_f]=X_{R(f)}.$
		
		Moreover, because $E_f=X_f+R,$ we know that 
		$$\iota_{[R,E_f]}\Omega=\iota_{[R,X_f]}\Omega=dR(f)-R(f)\theta-R(R(f))\eta,\quad \iota_{[R,E_f]}\eta=\iota_{[R,X_f]}\eta=0,$$
		thus $[R,E_f]=X_{R(f)}.$
	\end{proof}
	\begin{definition}
		We define the following set as the \textbf{strongest function set}:
		\[
		\mathcal S:=\{ H \in C^\infty(M) \mid R(H) = 0, \, \theta(X_H) = 0 \}.
		\]
		The elements of this set are referred to as \textbf{strongest functions}. 
		
		We define the following set as the \textbf{$R$-strong function set}:
	\[
	\tilde{\mathcal S}:=\{ H \in C^\infty(M) \mid R(H) = 0 \}.
	\]
	The elements of this set are referred to as \textbf{$R$-strong functions}.
		
		We define the following set as the \textbf{$\theta$-strong function set}:
		\[
		\mathcal S':=\{ H \in C^\infty(M) \mid   \theta(X_H) = 0 \}.
		\]
		The elements of this set are referred to as \textbf{$\theta$-strong functions}. 
	\end{definition}
	\begin{remark}
    For any $f \in \mathcal{S}'$, we observe that $R(f) \in \mathcal{S}'$ as well. In fact, by Proposition \ref{PR} and Remark \ref{Bao}, we know that
$$\theta(X_{R(f)})=\theta([R,X_f])=L_R(\theta(X_f))-(L_R\theta)(X_f)=0.$$
\end{remark}

	\begin{theorem}\label{The}
		We can prove that the COLCS bracket \( \{\cdot, \cdot \} \) defined above has the following properties:
		
		\begin{enumerate}
			\item \( \{g, f\} = X_f(g) - g \theta(X_f) = -X_g(f) + f \theta(X_g); \)
			\item \( X_{\{f, g\}} =[X_g, X_f]  \) and $(C^\infty(M),\{\cdot,\cdot\})$ is a  Lie algebra;
			%\item \( \mathcal{S}_E \) is a Lie algebra with respect to the bracket \( \{\cdot, \cdot\} \), but this bracket is not a Poisson bracket on this set. 
			\item \( \mathcal{S}, \tilde{\mathcal S} \) and \( \mathcal{S}' \) are all associative algebras under pointwise (function) multiplication$;$
			\item The bracket \( \{\cdot, \cdot\} \) is a Lie bracket on $\mathcal S,\tilde{\mathcal S}$ and $\mathcal S'$ respectively.
			\item The bracket \( \{\cdot, \cdot\} \) is a Poisson bracket on $\mathcal S$ and $\mathcal S'$ respectively.
			
		\end{enumerate}
	\end{theorem}
	
	\begin{proof}
		Firstly, we calculate
		\begin{align*}
			\{g, f\} &= \Omega(X_g, X_f) = \iota_{X_g} \Omega(X_f) = (dg - g \theta-R(g)\eta)(X_f) = X_f(g) - g \theta(X_f); \\
			\{g, f\} &= -\{f, g\} = -\Omega(X_f, X_g) = -\iota_{X_f} \Omega(X_g) = -(df - f \theta-R(f)\eta)(X_g) = -X_g(f) + f \theta(X_g).
		\end{align*}
		Moreover, we calculate
		\begin{align*}
			\iota_{[X_f, X_g]} \Omega &= L_{X_f}(\iota_{X_g} \Omega) - \iota_{X_g} L_{X_f} \Omega \\
			&= L_{X_f}(dg - g \theta-R(g)\eta) - \iota_{X_g} (\iota_{X_f} d\Omega + d \iota_{X_f} \Omega) \\
			&= d(X_f g) - (X_f g) \theta - g d(\theta(X_f)) - X_f(R(g))\eta-\iota_{X_g} (\theta(X_f) \Omega - \theta \wedge \iota_{X_f} \Omega - df \wedge \theta-dR(f)\wedge\eta) \\
			&= d(X_f g) - (X_f g) \theta - g d(\theta(X_f)) - X_f(R(g))\eta - \theta(X_f) \iota_{X_g} \Omega -R(f)\iota_{X_g}(\theta\wedge\eta)+\iota_{X_g}(d(R(f))\wedge\eta)\\
			&= d(X_f g) - (X_f g) \theta - g d(\theta(X_f))-X_f(R(g))\eta - \theta(X_f) dg + g \theta(X_f) \theta+R(g)\theta(X_f)\eta \\
			&\quad -R(f)\theta(X_g)\eta+X_g((R(f)))\eta,\\
			\iota_{X_{\{g, f\}}} \Omega &= d\{g, f\} - \{g, f\} \theta-R(\{g,f\})\eta \\
			&= d(X_f g - g \theta(X_f)) - (X_f(g) - g \theta(X_f)) \theta -R(X_f(g) - g \theta(X_f))\eta\\
			&= d(X_f g) - \theta(X_f) dg - g d(\theta(X_f)) -X_f(g) \theta + g \theta(X_f) \theta-R(X_f(g))\eta+R(g)\theta(X_f)\eta+gR(\theta(X_f))\eta,
		\end{align*}
		which means that
		\begin{align}\label{Eq3}
			\iota_{[X_f,X_g]-X_{\{g,f\}}}\Omega&=\iota_{[X_f, X_g]} \Omega  -\iota_{X_{\{g, f\}}} \Omega \nonumber\\
            &= -X_f(R(g))\eta+R(X_f(g))\eta-gR(\theta(X_f))\eta-R(f)\theta(X_g)\eta+X_g(Rf)\eta\nonumber\\
			&=([R,X_f](g)-gR(\theta(X_f))-R(f)\theta(X_g)+X_g(Rf))\eta\nonumber\\
			&=(X_{Rf}(g)-gR(\theta(X_f))-R(f)\theta(X_g)+X_g(Rf))\eta.
		\end{align}
Applying the vector field R to both sides of the above equation, we obtain
\begin{align*}
    X_{Rf}(g)-gR(\theta(X_f))-R(f)\theta(X_g)+X_g(Rf)=0.
\end{align*}
Moreover, we can calculate 
$$\iota_{[X_f,X_g]-X_{\{g,f\}}}\eta=\iota_{[X_f,X_g]}\eta=L_{X_f}(\iota_{X_g}\eta)-\iota_{X_g}L_{X_f}\eta=0,$$
which means that $[X_f,X_g]=X_{\{g,f\}}.$

Now, we prove the COLCS bracket is a Lie bracket on $C^\infty(M)$. 
		It's clear that the bracket \( \{\cdot, \cdot \} \) is a bilinear operation. Since \( \Omega \) has the skew-symmetry property, \( \{\cdot, \cdot \} \) also inherits this property. Next we prove that \( \{\cdot, \cdot \} \) satisfy the Jacobi identity:
		\begin{align}
			\{\{f, g\}, h\} + \{\{g, h\}, f\} + \{\{h, f\}, g\} = 0.
		\end{align}
		Firstly, let $f,g,h\in C^\infty(M)$, then 
		by $[X_f,X_g]=X_{\{g,f\}}$, we know that
		\begin{align*}
			\{h,\{f, g\}\} &= X_{\{f, g\}}(h)-h\theta(X_{\{f,g\}}) = [X_g, X_f](h)-h\theta([X_g,X_f]) \\
			&= X_g(X_f(h)) - X_f(X_g(h))-h\theta([X_g,X_f]) \\
			&= X_g(\{h, f\}+h\theta(X_f))  - X_f(\{h, g\}+h\theta(X_g))-hX_g(\theta(X_f))+hX_f(\theta(X_g)) \\
            &=X_g(\{h, f\})-X_f(\{h, g\})+X_g(h)\theta(X_f)-X_f(h)\theta(X_g)\\
            &=(X_g(\{h, f\})-X_f(h)\theta(X_g)+h\theta(X_f)\theta(X_g))-(X_f(\{h, g\})-X_g(h)\theta(X_f)+h\theta(X_f)\theta(X_g))\\
			&=\{\{h,f\},g\}-\{\{h,g\},f\}.
		\end{align*}
		which implies that \( \{\{f, g\}, h\} + \{\{g, h\}, f\} + \{\{h, f\}, g\} = 0 \). 
		Thus, we have proven that the $(C^\infty(M),\{\cdot,\cdot\})$ is a Lie algebra under the COLCS bracket.
        
		%By the discussion above, 
		%we know that if $f,g\in \mathcal S',$  we can calculate that 
		%\begin{align*}
			%\iota_{[X_f,X_g]-X_{\{g,f\}}}\Omega&=(X_{Rf}(g)-gR(\theta(X_f))-R(f)\theta(X_g)+X_g(Rf))\eta\\
			%&=(X_{Rf}(g)+X_g(Rf))\eta\\
			%&=(\{g,Rf\}+\{Rf,g\})\eta=0.
		%\end{align*} Moreover, we know that $\eta([X_f,X_g]-X_{\{g,f\}})=0,$ so $[X_f,X_g]=X_{\{g,f\}}$ for all $f,g\in\mathcal S'.$
		
		Next, we prove that  \( \mathcal{S},\tilde{\mathcal S} \) and \( \mathcal{S}' \) are all associative algebras under pointwise (function) multiplication. We just need to verify that 
		\begin{align*}
			R(fg)&=0,\quad \forall f,g\in\tilde{\mathcal S},\\
\theta(X_{fg})&=0,\quad \forall f,g\in\mathcal S',
		\end{align*}
		and 
		\begin{align*}
			\theta(X_{fg})=0,\quad R(fg)=0,\quad  \forall f,g\in\mathcal S.
		\end{align*}
        
        It is easy to see that for any $f,g\in\tilde{\mathcal S},$
        $$R(fg)=fR(g)+gR(f)=0.$$

		For $\forall f,g,h\in \mathcal S',$ we know that 
		\begin{align}
			\{gh, f\} &= X_f(gh) - gh \theta(X_f)= h X_f(g) + g X_f(h) \label{Lib1}\\
			h \{g, f\} + g \{h, f\} &= h X_f(g) - h g \theta(X_f) + g X_f(h) - gh \theta(X_f)=h X_f(g)  + g X_f(h) ,\label{Lib2}
		\end{align}
		so \( \{f, gh\} = h \{f, g\} + g \{f, h\} \). Moreover,
		\begin{align*}
			\iota_{X_{fg}}\Omega&=d(fg)-(fg)\theta-R(fg)\eta\\
			&=fdg+gdf-2(fg)\theta-fR(g)\eta-gR(f)\eta+(fg)\theta\\
			&=f\iota_{X_g}\Omega+g\iota_{X_{f}}\Omega+(fg)\theta.
		\end{align*}
		Define $W=X_{fg}-fX_g-gX_f$; the identity above gives $\iota_W\Omega=(fg)\theta$, and $\eta(W)=0$ since $\eta(X_H)=0$ for every Hamiltonian vector field. Hence $W$ takes values in $\xi=\ker\eta$.

		At any point where $fg\neq 0$, contracting both sides of $\iota_W\Omega=(fg)\theta$ with $W$ and using the definition of Hamiltonian vector field yields $(fg)\,\theta(W)=0$, so $\theta(W)=0$, hence $0=\theta(W)=\theta(X_{fg})-f\theta(X_g)-g\theta(X_f)=\theta(X_{fg})$.
		At any point $p$ where $f(p)g(p)=0$, we have $\iota_W\Omega\big|_p=0$ with $W|_p\in\xi_p$. Since $\Omega|_\xi$ is non-degenerate, $W|_p=0$. Because $f,g\in\mathcal S'$,
		$$\theta(X_{fg})=f\,\theta(X_g)+g\,\theta(X_f)+\theta(W)=\theta(W),$$
		so $\theta(X_{fg})(p)=0$.
		Therefore $\theta(X_{fg})=0$ everywhere, i.e.\ $fg\in\mathcal S'$. So \( \mathcal{S}' \) is an associative algebra under pointwise (function) multiplication. Similarly, we can get the same result for $\mathcal S$.
		
		Now, we prove the COLCS bracket is a Lie bracket on \( \mathcal S,\tilde{\mathcal S} \) and $\mathcal S'$ respectively. 
To verify that the bracket \( \{\cdot, \cdot\} \) is a Lie bracket on $\tilde{\mathcal S},$ it suffices to show that $R(\{f,g\})=0$ for any $f,g\in\tilde{\mathcal{S}}$. Indeed, using Proposition~\ref{PR} and Remark~\ref{Bao},
\begin{align*}
    R(\{f,g\}) &= (L_R\Omega)(X_f,X_g) + \Omega([R,X_f],X_g) + \Omega(X_f,[R,X_g]) \\
               &= \Omega(X_{R(f)},X_g) + \Omega(X_f,X_{R(g)}) = 0.
\end{align*}
 We continue to verify that $\mathcal{S}'$ also carries a Lie algebra structure. We only need to prove that $$\theta(X_{\{f,g\}})=0,\quad \forall f,g\in\mathcal S'.$$ By the statement above, we know that for any $f,g\in\mathcal S'$,
 $$\theta(X_{\{f, g\}})=\theta([X_g, X_f])=L_{X_g}(\theta(X_f))-(L_{X_g}\theta)(X_f)=0, $$
which means that the bracket \( \{\cdot, \cdot\} \) is a Lie bracket on $\mathcal S'.$

Furthermore, relations \eqref{Lib1} and \eqref{Lib2} imply that the COLCS bracket satisfies the Leibniz rule on both $\mathcal{S}$ and $\mathcal{S}'$. Consequently, this bracket does define a Poisson structure on $\mathcal{S}$ and $\mathcal{S}'$.
	\end{proof}
	\begin{remark}
		According to Theorem~\ref{The}, it follows that $\mathcal{S}$ is a Lie subalgebra of $\mathcal{S}'$ and $\tilde{\mathcal S}$.
	\end{remark}
	
\section{Lie Integrability of Hamiltonian systems on COLCS manifold}\label{lie-integrability}
	As noted in Remark~\ref{R:scope-integrability}, the integrability results in this section concern the Hamiltonian vector field $X_H$ (and its restrictions to appropriate invariant submanifolds), not the evolution vector field $E_H$.
	A classical theorem due to Sophus Lie (commonly called Lie's theorem \cite{Kozlov2}) states that a differential equation can be integrated by quadratures provided its associated vector field possesses sufficiently many symmetries forming a solvable Lie algebra under the standard Lie bracket.
For notational simplicity, the family of brackets introduced in the preceding section on functions over manifolds will be collectively denoted as the Poisson bracket hereafter.
	\begin{definition}[Lie Algebra]
		Let \(\mathfrak{g}\) be a vector space equipped with a bilinear operation \([\cdot, \cdot]: \mathfrak{g} \times \mathfrak{g} \to \mathfrak{g}\), called the \emph{Lie bracket}, satisfying the following conditions:
		\begin{enumerate}
			\item \textbf{Antisymmetry:} For all \(X, Y \in \mathfrak{g}\), we have \([X, Y] = -[Y, X]\).
			\item \textbf{Jacobi identity:} For all \(X, Y, Z \in \mathfrak{g}\),
			\[
			[X, [Y, Z]] + [Y, [Z, X]] + [Z, [X, Y]] = 0.
			\]
			\item \textbf{Bilinearity:} The Lie bracket is linear in both arguments.
		\end{enumerate}
	\end{definition}
	
	\begin{definition}[Lie Subalgebra]
		Let \(\mathfrak{g}\) be a Lie algebra. A \emph{Lie subalgebra} of \(\mathfrak{g}\) is a vector subspace \(L \subset \mathfrak{g}\) such that
		\[
		[X, Y] \in L, \quad \forall X, Y \in L.
		\]
	\end{definition}
	
	\begin{definition}[Ideal of a Lie Algebra]
		Let \(\mathfrak{g}\) be a Lie algebra. A vector subspace \(W \subset \mathfrak{g}\) is called an \emph{ideal} of \(\mathfrak{g}\) if
		\[
		[X, Y] \in W, \quad \forall X \in W,\, Y \in \mathfrak{g}.
		\]
	\end{definition}
	\begin{remark}
		Every ideal of a Lie algebra \(\mathfrak{g}\) is also a Lie subalgebra of \(\mathfrak{g}\).
	\end{remark} 
	
	\begin{definition}[Solvable Lie Algebra]
		Let \(\mathfrak{g}\) be a Lie algebra. We say that \(\mathfrak{g}\) is solvable if there exists a chain of Lie subalgebras, let us say $L_1,...,L_n$, such that
		\[
		\{0\} = L_0 \subset L_1 \subset L_2 \subset \cdots \subset L_{n-1} \subset L_n = \mathfrak{g}, \quad \text{with } n = \dim(\mathfrak{g}),
		\]
		such that each \(L_i\) is an ideal in \(L_{i+1}\) with codimension one, for \(i = 0,1,\dots,n-1\).
	\end{definition}
	\begin{definition}
		We say that a vector field $X$ is a symmetry of the vector field $Y$ if $[X,Y]=0,$ where $[\cdot,\cdot]$ is the Lie bracket of vector fields.
	\end{definition}
	\begin{theorem} \cite[Lie Integrability Theorem]{Arnold2}\label{TArnold}
		Let \(X_1, \dots, X_n\) be linearly independent smooth vector fields on $n$ dimensional manifold \(M\), generating a solvable Lie algebra \(\mathfrak{g}\). Suppose that one of the vector fields (denoted \(X_1\)) defines a dynamical system and that all vector fields are symmetries of \(X_1\), i.e., 
		\[
		[X_1, X_i] = 0, \quad \text{for } i = 2, \dots, n.
		\]
		Then the dynamical system defined by $X_1$ 
		can be solved by quadratures.
	\end{theorem}
In the geometric study of integrable systems, a fundamental structure is that of a torus bundle. Recall that a smooth map $\pi: E \to B$ between manifolds is called a $\mathbb{T}^n$-bundle if its fibers are diffeomorphic to the $n$-dimensional torus $\mathbb{T}^n \cong \mathbb{R}^n / \mathbb{Z}^n$, and the structure admits a local trivialization. Specifically, there exists an open covering $\{U_\alpha\}$ of the base $B$ and diffeomorphisms $\Phi_\alpha: \pi^{-1}(U_\alpha) \to U_\alpha \times \mathbb{T}^n$ such that the transition functions act on the fiber $\mathbb{T}^n$ by translations. 
            
In the context of integrable systems, this structure arises naturally from the level sets of functionally independent first integrals: if the level sets are compact and connected, the Arnold-Liouville theorem (and its non-Hamiltonian generalizations) guarantees that the phase space foliates into invariant tori, giving the manifold the structure of a $\mathbb{T}^n$-bundle over the space of conserved quantities.

% \begin{definition}[\(\mathbb{T}^n\)-Bundle]
% Let \(E\) and \(B\) be smooth manifolds. A smooth map \(\pi: E \to B\) is called a \(\mathbb{T}^n\)-bundle, or an \(n\)-torus bundle, if its fibers are diffeomorphic to the \(n\)-dimensional torus \(\mathbb{T}^n \cong \mathbb{R}^n / \mathbb{Z}^n\). Furthermore, the bundle must admit a local trivialization. That is, there exists an open covering \(\{U_\alpha\}_{\alpha \in I}\) of the base space \(B\) and a collection of diffeomorphisms
% \[
% \Phi_\alpha: \pi^{-1}(U_\alpha) \longrightarrow U_\alpha \times \mathbb{T}^n,
% \]
% such that \(\operatorname{pr}_1 \circ \Phi_\alpha = \pi\), where \(\operatorname{pr}_1\) is the projection onto the first factor. For any two overlapping charts \(U_\alpha\) and \(U_\beta\), the transition functions are given by
% \[
% \Phi_\beta \circ \Phi_\alpha^{-1}: (U_\alpha \cap U_\beta) \times \mathbb{T}^n \longrightarrow (U_\alpha \cap U_\beta) \times \mathbb{T}^n,
% \]
% which take the form \((x, \theta) \mapsto (x, g_{\beta\alpha}(x)(\theta))\). The crucial structural condition is that the maps \(g_{\beta\alpha}(x): \mathbb{T}^n \to \mathbb{T}^n\) act on the fiber \(\mathbb{T}^n\) by \emph{translations}.
% \end{definition}

% \begin{remark}
% The condition that the transition functions act by translations implies that the structure group of the bundle reduces to the abelian group \(\mathbb{T}^n\) itself, acting on itself by left (or right) multiplication. This is a significant constraint that distinguishes \(\mathbb{T}^n\)-bundles from general fiber bundles with torus fibers.
% \end{remark}
    
	Now, we can introduce another theorem.

	\begin{theorem}\cite{Bogoyavlenskij,Zung} \label{TBogo}
		Assume that on a manifold \( M \) there exist:
		\begin{enumerate}
			\item A submersion \( F = (F_1, \dots, F_k) : M \rightarrow \mathbb{R}^k \) with compact and connected fibers, for some \( 1 \leq k < \dim M \).
			
			\item A collection of \( n = \dim M - k \) vector fields \( Y_1, \dots, Y_n \) that are everywhere linearly independent, pairwise commuting, and tangent to the fibers of \( F \), i.e.,
			\[
			[Y_i, Y_j] = 0, \quad \mathcal{L}_{Y_i} F_r = 0, \quad \forall i,j = 1, \dots, n, \; r = 1, \dots, k.
			\]
		\end{enumerate}
		Then:
		\begin{itemize}
			\item[i.] The map \( F : M \rightarrow F(M) \subset \mathbb{R}^k \) defines a \( \mathbb{T}^n \)-bundle.
			
			\item[ii.] Any vector field \( X \) on \( M \) satisfying
			\[
			\mathcal{L}_X F_r = 0 \quad \text{and} \quad [X, Y_i] = 0, \quad \forall i = 1, \dots, n, \; r = 1, \dots, k,
			\]
			is conjugate to a constant vector field on \( \mathbb{T}^n \) via each bundle chart of \( F : M \rightarrow F(M) \).
		\end{itemize}
	\end{theorem}
	The above notion of integrability is often termed \textbf{non-Hamiltonian integrability}, and the tuple $(Y_1,\dots,Y_n,F_1,\dots,F_k)$ is called an \textbf{integrable system of type $(n,k)$} on $M$. This terminology does not imply the absence of a Hamiltonian structure; rather, it indicates that we disregard any such structure and focus solely on the commuting flows and first integrals in the definition. In fact, a Hamiltonian system on a symplectic manifold that is integrable in the Liouville--Arnold sense is also integrable in this broader sense.

Observe that the vector fields $Y_1,\dots,Y_n$ are tangent to the fibers of the map $F=(F_1,\dots,F_k)$. The system is said to be \textbf{regular} on a fiber $L$ of $F$ if both $Y_1\wedge\cdots\wedge Y_n\neq 0$ and $dF_1\wedge\cdots\wedge dF_k\neq 0$ hold everywhere on $L$. Moreover, the system is called \textbf{proper} if the map $F\colon M\to\mathbb{R}^k$ is topologically proper (so every fiber is compact) and regularity holds on almost every fiber.

With these preliminaries, we can now formulate a Lie-type theorem for $\theta$-strong Hamiltonian vector fields on a COLCS manifold.
	\begin{theorem}\label{TZ}
		Let \((M,\Omega,\theta,\eta)\) be a \((2n+1)\)-dimensional COLCS manifold and let \(X_H=X_{f_1}\) be a $\theta$-strong Hamiltonian vector field with $\theta$-strong Hamiltonian function \(H=f_1\in\mathcal S'\). Suppose there exist functionally independent \(2n - k\) functions \(f_2, \ldots, f_{2n-k+1}\) that are constants of motion for \(X_{f_1}\) and satisfy the following conditions:
		\begin{enumerate}
			\item For  \(2 \leq i \leq k\leq n\), the  Hamiltonian vector field \(X_{f_i}\) associated with \(f_i\) is a $\theta$-strong Hamiltonian vector field, i.e.,
			\[
	\iota_{X_{f_i}}\Omega=df_i-f_i\theta-R(f_i)\eta,\quad \theta(X_{f_i})= 0, \quad i = 2, \ldots, k;
			\]

			\item $f_2,...,f_{k}$ satisfy	$$	\{f_i, f_j\} =\sum_{l=1}^{k} c_{ij}^l f_l, \quad \text{with } c_{ij}^l \in \mathbb{R}, \quad i, j = 2, \ldots, k; $$
            and
            $$ \{f_i,f_j\}=0,\quad 2\leq i\leq k,\quad k+2\leq j\leq 2n-k+1;$$
			
			\item The functions \(f_1, \ldots, f_{k}\) generate a solvable Lie algebra under the Poisson bracket;
			
			\item On the level set
			\[
			M_{f} = \left\{ x \in M : f_i(x) = c_i,\ c_i \in \mathbb{R},\ i = 1, \ldots, 2n-k+1 \right\},
			\] the constants satisfy
			\[
				\sum_{l=1}^{k}c_{ij}^l c_l = 0, \quad \text{for all } i, j = 2, \ldots, k.
			\]
		\end{enumerate}
		Then \(M_{f}\) is a smooth submanifold of \(M\) of dimension \(k\), and the solutions of the $\theta$-strong Hamiltonian vector field \(X_{f_1}\) restricted to \(M_{f}\) can be obtained by quadratures. 
	\end{theorem}
%     Before presenting the proof, we briefly recall the notion of functional independence
% 	and some of its implications. Let \(N\) be a smooth manifold. A collection of functions \(F_1,\dots,F_k\), with \(k \le \dim N\), is said to be functionally independent at \(p \in N\) if their differentials \(dF_1|_p,\dots,dF_k|_p\) are linearly independent. Equivalently, \(p\) is a regular point of the map \(F = (F_1,\dots,F_k): N \to \mathbb R^k\).

% Define the level set
% \[
% N_f = \{ x \in N \mid F_i(x)=c_i,\; c_i \in \mathbb R \}.
% \]
% If \(F_1,\dots,F_k\) are functionally independent at every point of \(N_f\), then the regular level set theorem \cite{Lee} guarantees that \(N_f\) is an embedded smooth submanifold of \(N\) with dimension \(\dim N - k\).
	\begin{proof}
		The functional independence of \(f_1, \ldots, f_{2n-k+1}\) on \(
M_f = \{ x \in M \mid f_i(x)=c_i,\; c_i \in \mathbb R \}
\) implies that \(M_{f}\) is a smooth submanifold of dimension \(2n+1 - (2n-k+1) = k\).
		As noted in equation \eqref{Eq3}, for any Hamiltonian functions \(f\), \(g\in C^\infty(M)\), we have the identity
		\begin{equation}\label{fg2}
			X_{\{f, g\}} = [X_g, X_f].
		\end{equation}
		Applying this to our setting, we obtain
		\[
		X_{\{f_i, f_j\}} = [X_{f_i}, X_{f_j}], \quad \text{for all } i, j = 1, \ldots, k.
		\]
	Therefore, the vector fields \(X_{f_1}, \ldots, X_{f_{k}}\) generate a solvable Lie algebra under the Lie bracket of vector fields.
		
		Next, observe that each \(X_{f_i}\) is tangent to \(M_{f}\). Indeed, by Theorem \ref{The}, for \(i, j = 1, \ldots, k\),
		\begin{align}\label{Xf}
				(X_{f_i} f_j)|_{M_{f}} = (\{f_j, f_i\}+f_j\theta(X_{f_i}))|_{M_{f}}= \sum_{l=1}^{k}c_{ij}^l f_l|_{M_{f}} = \sum_{l=1}^{k}c_{ij}^l c_l = 0.
		\end{align}
		
		Therefore, the flow of \(X_{f_1}\) remains within \(M_{f}\), and the dynamics restricted to \(M_{f}\) are governed by the differential equation
		\[
		\dot{x} = X_{f_1}(x),
		\]
		where \((x^1, \ldots, x^k)\) are local coordinates on \(M_{f}\).
		
		Since \(X_{f_1}\) belongs to a solvable Lie algebra generated by \(X_{f_1}, \ldots, X_{f_{k}}\)	
		the system is solvable by quadratures. That is, its solutions can be obtained through successive integrations.
	\end{proof}
   \begin{example}
Consider a COLCS manifold defined by 
\[
(M = \mathbb R^5,\; \Omega = e^x (dx \wedge dy + du \wedge dv),\; \theta = dx,\; \eta = dt),
\]
where $\mathbb R^5$ is equipped with coordinates $(x, y, u, v, t)$. 
Let $f_1 = H = tu$. A direct computation yields the Hamiltonian vector field associated with $f_1$:
\[
X_{f_1} = X_H = tu e^{-x} \frac{\partial}{\partial y} - t e^{-x} \frac{\partial}{\partial v}.
\]
Now set $f_2 = u$ and $f_3 = x$. Then one finds 
\[
X_{f_2} = u \frac{\partial}{\partial y} - \frac{\partial}{\partial v},
\]
and moreover $\theta(X_{f_1}) = \theta(X_{f_2}) = 0$. 
Using the definition of the bracket in this setting, we obtain the following pairwise brackets:
\begin{align*}
\{f_2, f_1\} &= X_{f_1}(f_2) - f_2 \theta(X_{f_1}) \\
&= tu e^{-x} \frac{\partial u}{\partial y} - t e^{-x} \frac{\partial u}{\partial v} - u \, dx\!\left( tu e^{-x} \frac{\partial}{\partial y} - t e^{-x} \frac{\partial}{\partial v} \right) = 0,\\[4pt]
\{f_3, f_1\} &= X_{f_1}(f_3) - f_3 \theta(X_{f_1}) \\
&= tu e^{-x} \frac{\partial x}{\partial y} - t e^{-x} \frac{\partial u}{\partial x} - x \, dx\!\left( tu e^{-x} \frac{\partial}{\partial y} - t e^{-x} \frac{\partial}{\partial v} \right) = 0,\\[4pt]
\{f_3, f_2\} &= X_{f_2}(f_3) - f_3 \theta(X_{f_2}) \\
&= u \frac{\partial x}{\partial y} - \frac{\partial x}{\partial v} - x \, dx\!\left( u \frac{\partial}{\partial y} - \frac{\partial}{\partial v} \right) = 0.
\end{align*}
All these brackets vanish, indicating that the functions $f_1, f_2, f_3$ are in involution. Consequently, by Theorem~\ref{TZ}, on the common level set 
\[
M_f = \bigl\{ x \in \mathbb R^5 : f_i(x) = c_i,\ i = 1,2,3 \bigr\},
\]
the Hamiltonian vector field $X_{f_1}$ is integrable by quadratures. In other words, the solutions of the corresponding differential equations can be obtained by performing a finite number of algebraic operations and integrations of known functions.
\end{example}
	%Here, we assume that \( H \in \mathcal{S}' \). In what follows, we remove this restriction and present another integrability theorem.

\section{Scaling Symmetries and First Integrals}\label{scaling-symmetries}
%%%%%%%%%%%%%%%%%%%%%%%%%%%%%%%%%%%%%%%%%%%%%%%%%%

Having established Lie integrability scheme for Hamiltonian systems on COLCS manifolds, we now turn to another structural aspect of such systems: symmetries that scale the geometric and dynamical data. In \cite{Azuaje}, the authors introduced the concept of a scaling symmetry of degree $(\Lambda,\beta)\in\mathbb{R}^2$ for time-dependent Hamiltonian systems. In the present context, we analogously introduce a notion of scaling symmetry of degree $(\Lambda,\beta,\gamma)\in\mathbb{R}^3$ for Hamiltonian systems defined on COLCS manifolds that may depend on an external parameter.

As will be shown, these symmetries play a crucial role in generating conservation laws and revealing hidden integrability properties \cite{Grabowska,Haller1998,Kozlov,Mezic1994,Zhao4,Zhao5}. In particular, they provide a systematic method for constructing new conservation laws from known ones, for establishing relationships between different scaling symmetries, and for relating distinct Hamiltonian systems on the same COLCS manifold. This section introduces the notion of a scaling symmetry of degree $(\Lambda,\beta,\gamma)$ and investigates its algebraic structure and dynamical consequences.
		\begin{definition}\label{D5}
		For a COLCS manifold $(M,\Omega,\theta,\eta)$, a vector field $X\in\mathfrak{X}( M)$ shall be called a scaling symmetry of degree $(\Lambda,\beta,\gamma)\in\mathbb{R}^3$  for the Hamiltonian system $( M,{\Omega},{\theta},\eta,H)$ when
		\begin{enumerate}
			\item [i)] $L_X\eta=\gamma \eta,$ 
			\item[ii)] $[X,R]=-\gamma R,$
			\item[iii)] $L_{X}{\Omega}=\beta{\Omega}$,
			\item [iv)] $L_{X}{\theta}=0$, (therefore $\theta(X)$ is constant)
			\item [v)] $L_{X}H=\Lambda H$.
		\end{enumerate}
	 \end{definition}
	% \begin{remark}\label{R5}
	% 	We can see that if $[X,R]=aR,$ for some constant $a\in\mathbb R,$ then condition $i)$ is equivalent to condition $a=-\gamma.$ In fact,
	% 	$$0=L_X(1)=L_X(i_R\eta)=i_{L_XR}\eta+i_RL_X\eta=ai_R\eta+\gamma i_R\eta =a+\gamma,$$
	% 	which means that $a=-\gamma.$
	% \end{remark}

	% 	However, without the condition $[X,R]=aR$ with $a\in\mathbb{R}$, we cannot obtain the conclusion described in Remark~\ref{R5}. That is, if $[X,R]\neq aR$ for any $a\in\mathbb{R}$, then $L_X\eta=\gamma\eta$ does not imply $[X,R]=-\gamma R$.
        
    %     \notepan{I don't get this sentence at all}
    %     \notezhao{I have rephrased it in a different way.}
	% 	\notepan{It still sounds a bit weird to me.}

		\begin{remark}\label{R5}
			Conditions (i) and (ii) in Definition \ref{D5} are closely related. Recall that for the Reeb vector field, $i_R\eta = 1$. Applying the Lie derivative along $X$ yields:
			$$0 = L_X(i_R\eta) = i_{[X,R]}\eta + i_R(L_X\eta).$$
			If condition (i) holds, this equation simplifies to $i_{[X,R]}\eta = -\gamma$. Geometrically, this implies that condition (i) fixes the Reeb component of the commutator to be precisely $-\gamma R$. Therefore, condition (ii) is equivalent to imposing condition (i) alongside the additional constraint that $[X,R]$ has no projection onto $\ker(\eta)$.
		\end{remark}

		\begin{example}
		The COLCS manifold $(M, \Omega, \theta', \eta)$ in Example \ref{E1} is described in local coordinates $(x_1, \dots, x_{2n}, t)$.
		Consider the vector field $X = t \frac{\partial}{\partial x_1}$.
		It is clear that $R = \frac{\partial}{\partial t}$ and
		$$[X, R] = -\frac{\partial}{\partial x_1} \quad \text{and} \quad L_X\eta = L_X(dt) = 0.$$
		\end{example}
	We now provide another example to demonstrate that the symmetry in Definition \ref{D5} indeed exists.
	\begin{example}\label{E3}
		Let \( (\mathbb R\times T^*Q, \Omega, \theta,\eta=dt) \) be a COLCS  manifold, where
		\[
	\Omega = dp_1 \wedge dq_1 + dp_2 \wedge dq_2 - p_2 q_1\,dq_1 \wedge dq_2, \quad Q = \mathbb{R}^2\setminus\{(0,0)\} , \quad \theta = q_1\,dq_1.
		\]
	 	Consider the Hamiltonian \( H = C (p_1^2+q_2^2+t^2) \), where \( C \in \mathbb{R} \), and the vector field \( X = p_1\frac{\partial}{\partial p_1}+q_2\frac{\partial}{\partial q_2}+t\frac{\partial}{\partial t} \).  Moreover, we can compute:
		\[
		L_X \Omega = \Omega, \qquad L_X \theta = 0, \qquad L_X H =  2H, \qquad L_X\eta=\eta, \qquad[X,R]=[X,\frac{\partial}{\partial t}]=- \frac{\partial}{\partial t}= -R.
		\]
		Thus,  \( X = p_1\frac{\partial}{\partial p_1}+q_2\frac{\partial}{\partial q_2}+t\frac{\partial}{\partial t} \)  is a scaling symmetry of degree $(2,1,1)$.
	\end{example}
    According to our definition, different scaling symmetries can, in certain cases, be transformed into one another.
\begin{proposition}
    If $X$ is a scaling symmetry of degree $(\Lambda,\beta,\gamma)$ for the Hamiltonian system $(M,\Omega,\theta,\eta,H)$, then

    i) If $\gamma \neq 0$, then we can construct a scaling symmetry of degree $(\frac{\Lambda}{\gamma},\frac{\beta}{\gamma},1)$;

    ii) If $\beta \neq 0$, then we can construct a scaling symmetry of degree $(\frac{\Lambda}{\beta},1,\frac{\gamma}{\beta})$;

    iii) for any $\alpha\in\mathbb R,\alpha X$ is a scaling symmetry of degree $(\alpha\Lambda,\alpha\beta,\alpha\gamma)$.
\end{proposition}

\begin{proof}
    For case i), let $\tilde{X} = \frac{1}{\gamma} X$. A direct computation yields
    \[
    L_{\tilde{X}}\eta = \eta,\quad [\tilde{X}, R] = -R,\quad L_{\tilde{X}}\Omega = \frac{\beta}{\gamma}\,\Omega,\quad L_{\tilde{X}}\theta = 0,\quad L_{\tilde{X}}H = \frac{\Lambda}{\gamma} H.
    \]

    For case ii), let $\hat{X} = \frac{1}{\beta} X$. Then we obtain
    \[
    L_{\hat{X}}\eta =\frac{\gamma}{\beta} \,\eta,\quad [\hat{X}, R] = -\frac{\gamma}{\beta}R,\quad L_{\hat{X}}\Omega = \Omega,\quad L_{\hat{X}}\theta = 0,\quad L_{\hat{X}}H = \frac{\Lambda}{\beta} H.
    \]

    For case iii), let $\hat{X} = \alpha X$. Then we can calculate

    $$ L_{\hat{X}}\eta =\alpha\gamma\,\eta,\quad [\hat{X}, R] = -\alpha\gamma R,\quad L_{\hat{X}}\Omega = \alpha\Omega,\quad L_{\hat{X}}\theta = 0,\quad L_{\hat{X}}H = \alpha\Lambda H.$$
\end{proof}
	We will see that a scaling symmetry of degree $(\Lambda,\beta,\gamma)$ is neither necessarily a standard symmetry (i.e., $[X, X_H] = 0$) nor a Hamiltonian vector field. In fact, a scaling symmetry of degree $(\Lambda,\beta,\gamma)$ rescales the Hamiltonian dynamics by a constant factor as stated in the following proposition.
	 \begin{proposition}
	 	If $X$ is a scaling symmetry of degree $(\Lambda,\beta,\gamma)$  for the Hamiltonian system $(M,\Omega,\theta,\eta,H)$, then $L_{X}X_{H}$ is a Hamiltonian vector field with Hamiltonian function $(\Lambda-\beta)H$, and $L_{X}X_{H}=(\Lambda-\beta)X_{H}$.
	 \end{proposition}
	 \begin{proof}
	 	Let us assume that  $X$ is a scaling symmetry of degree $(\Lambda,\beta,\gamma)$ for $(M,\Omega,\theta,\eta,H)$, we have
        
	 	\begin{equation}
	 		\begin{split}
	 			i_{(L_{X}X_{H})}\Omega=i_{[X,X_{H}]}\Omega&=L_{X}(i_{X_{H} }\Omega)-i_{X_{H}} L_{X}\Omega\\
	 			&= L_{X}(dH-H\theta-(i_RdH)\eta)-\beta i_{X_{H}}\Omega\\			&=dL_{X}H-L_XH\theta-HL_{X}\theta-(i_{[X,R]}dH)\eta\\
	 			&\quad-(i_RdL_XH)\eta-R(H)L_X\eta-\beta d^{\theta}H+\beta R(H)\eta\\
	 			&=d(\Lambda H)-(\Lambda H)\theta-R(\Lambda H)\eta- \beta d^\theta H+\beta R(H)\eta\\
	 			&= d^{\theta}(\Lambda H)- d^{\theta}(\beta H)-R(\Lambda H-\beta H)\eta\\
	 			&= d^{\theta}((\Lambda-\beta)H)-R((\Lambda-\beta)H)\eta,
	 		\end{split}
	 	\end{equation}
	 	we can also see that $$i_{L_XX_H}\eta=L_X(i_{X_H}\eta)-i_{X_H}L_X\eta=-\gamma i_{X_H}\eta=0.$$
	 	So $L_{X}X_{H}$ is a Hamiltonian vector field with Hamiltonian function $(\Lambda-\beta)H$, i.e.,
	 	\begin{equation}
	 		L_{X}X_{H}=X_{(\Lambda-\beta)H}.
	 	\end{equation}
	 	Moreover, by $$d^{\theta}((\Lambda-\beta)H)-R((\Lambda-\beta)H)\eta=(\Lambda-\beta)(d^\theta H-R(H)\eta)=(\Lambda-\beta)i_{X_H}\Omega=i_{(\Lambda-\beta)X_H}\Omega,$$ we know that 
	 	$$L_XX_H=(\Lambda-\beta)X_H.$$
	 \end{proof}
     \begin{remark}
         According to the discussion above, we get that  $X$ is a $\lambda$-symmetry or a special $\sigma$-symmetry of the Hamiltonian vector field $X_H$ (see \cite{Zhao}). Specially,
         when $\Lambda=\beta,$ $X$ is a symmetry of the Hamiltonian vector field $X_H,$ i.e., $[X,X_H]=0$.
     \end{remark}
     \begin{proposition}
      We can also find that as long as we have a first integral $G$ and a scaling symmetry  of degree $(\Lambda, \beta, \gamma)$ $X$ of the Hamiltonian vector field $X_H$, we can obtain a family of first integrals $L_X^n G$, $n = 0, 1, 2, \dots$, where $L^0_XG=G$.
     \end{proposition}
    \begin{proof}
    In fact, we can compute that
    $$L_{X_H} L_X G = L_{[X_H, X]} G - L_X L_{X_H} G = (\beta - \Lambda) L_{X_H} G = 0.$$
    By induction, we obtain that for any $n > 1$,
    $$L_{X_H} L_X^n G = 0.$$
\end{proof}
	 Now consider a special case of a scaling symmetry of degree \( (\Lambda, \beta,\gamma) \), namely the case \( (\Lambda, \beta,\gamma) = (0, 0,0) \). This means that the vector field \( X \) satisfies
	 \[
	 L_X \Omega = 0, \qquad L_X \theta = 0, \qquad L_X H = 0, \qquad L_X\eta=0, \qquad[X,R]=0.
	 \]
	 In the following, we provide an example to demonstrate that such symmetries do indeed exist.
	 \begin{example}
	 	Let $(M,\Omega,\theta,\eta)$ be the COLCS manifold given by Example \ref{E3}, let $H=e^{\Lambda q_1}$ and $X=\frac{\partial}{\partial q_2}+\frac{\partial}{\partial t}$, we can see that  \[
	 	L_X \Omega = 0, \qquad L_X \theta = 0, \qquad L_X H = 0,\qquad L_X\eta=0, \qquad[X,R]=0.
	 	\]
	 	Thus, $X=\frac{\partial}{\partial q_2}+\frac{\partial}{\partial t}$ is a scaling symmetry of degree $(0,0,0)$.
	 \end{example}
	 \begin{remark}
	 	It is easy to see that all the scaling symmetry of degree $(0,0,0)$ preserve the volume form $\Omega^n\wedge\eta$.
	 \end{remark}
	In \cite{Azuaje}, the authors also established an interesting property concerning scaling symmetries of degree \( (\Lambda, \beta) \), namely that ``the set of all scaling symmetries of degree \( (\Lambda, \beta) \) forms a Lie algebra, and the Lie bracket of any two such symmetries is a scaling symmetry of degree \( (0, 0) \)". We now show that this property also holds for the scaling symmetries of degree \( (\Lambda, \beta, \gamma) \) introduced in this paper.
	 \begin{proposition}
	 	The set of all scaling symmetries of degree \( (\Lambda, \beta,\gamma) \) forms a Lie algebra. Moreover, the Lie bracket of any two scaling symmetries of degree \( (\Lambda, \beta,\gamma) \) is a scaling symmetry of degree \( (0, 0,0) \).
	 \end{proposition}
	 \begin{proof}
	 	In fact, it suffices to show that the Lie bracket of any two scaling symmetries of degree \( (\Lambda, \beta,\gamma) \) is another scaling symmetry of degree \( (0, 0,0) \). 
	 	
	 	Let \( X_1 \) and \( X_2 \) be scaling symmetries of degrees \( (\Lambda_1, \beta_1,\gamma_1) \) and \( (\Lambda_2, \beta_2,\gamma_2) \), respectively. Using the identity 
	 	\[
	 	L_{[X_1, X_2]} = L_{X_1} L_{X_2} - L_{X_2} L_{X_1},
	 	\]
	 	we compute:
	 	\begin{align*}
	 		L_{[X_1, X_2]} \Omega 
	 		&= L_{X_1} L_{X_2} \Omega - L_{X_2} L_{X_1} \Omega 
	 		= \beta_2 L_{X_1} \Omega - \beta_1 L_{X_2} \Omega 
	 		= \beta_2 \beta_1 \Omega - \beta_1 \beta_2 \Omega = 0, \\
	 		L_{[X_1, X_2]} \theta 
	 		&= L_{X_1} L_{X_2} \theta - L_{X_2} L_{X_1} \theta = 0, \\
	 		L_{[X_1, X_2]} H 
	 		&= L_{X_1} L_{X_2} H - L_{X_2} L_{X_1} H 
	 		= \Lambda_2 L_{X_1} H - \Lambda_1 L_{X_2} H 
	 		= \Lambda_2 \Lambda_1 H - \Lambda_1 \Lambda_2 H = 0,\\
	 		L_{[X_1,X_2]}\eta&=L_{X_1}L_{X_2}\eta-L_{X_2}L_{X_1}\eta=\gamma_2L_{X_1}\eta-\gamma_1L_{X_2}\eta=\gamma_2\gamma_1\eta-\gamma_1\gamma_2\eta=0,\\
	 		[[X_1,X_2],R]&=	L_{[X_1,X_2]}R=L_{X_1}L_{X_2}R-L_{X_2}L_{X_1}R=-\gamma_2L_{X_1}R+\gamma_1L_{X_2}R=\gamma_2\gamma_1R-\gamma_1\gamma_2R=0.
	 	\end{align*}
	 	Therefore, \( [X_1, X_2] \) is a scaling symmetry of degree \( (0, 0,0) \), which completes the proof.
	 \end{proof}
Here, we give another interesting property about scaling symmetries.
	 \begin{proposition}
	 	If Hamiltonian system $(M, \Omega,\theta, \eta, H)$ admits a scaling symmetry $X$ of degree $(\Lambda,\beta\neq\theta(X),\gamma\neq 0)$, then there exists a 1-form $\omega$  and a function $\Theta$ such that $$d^\theta\omega=\Omega,\quad d\Theta = \eta.$$ 
	 \end{proposition}
	 \begin{proof}
	 	  Let $X$ be a scaling symmetry of degree $(\Lambda,\beta,\gamma)$ for $(M, \Omega,\theta, \eta, H)$.  
	 	Define $\tilde\omega = i_X  \Omega$ and $\tilde\Theta=i_X\eta$. Then   \[
	 	d^\theta \tilde\omega = d^\theta(i_X  \Omega) =d^\theta(i_X  \Omega)+\theta(X)  \Omega-\theta(X)  \Omega  = L_X\Omega-\theta(X)  \Omega =(\beta-\theta(X) )\Omega,
	 	\]
	 	since $\beta\neq\theta(X),$ let $\omega=\frac{\tilde\omega}{\beta-\theta(X)}$, then we know that $d^\theta\omega=\Omega.$ Moreover, by $d\eta=0,$ we can calculate
	 	$$d\tilde\Theta=di_X\eta=di_X\eta+i_Xd\eta=L_X\eta=\gamma\eta.$$
	 Because $\gamma\neq 0,$ let $\Theta=\frac{\tilde\Theta}{\gamma}$, we know that $d\Theta=\eta.$	 
	 \end{proof}
	 	\section*{Acknowledgment} The research of author is supported by NSFC (Grant No. 12401234).
 %The author expresses his deep gratitude to anonymous referees for their valuable comments which have improved the paper.
	$\\$
	
	\noindent$\mathbf{Conflict\;of\;interest\;statement.}$ On behalf of all authors, the corresponding author states that there is no conflict of interest.
	
	$\\$
	\noindent$\mathbf{Data\;availability.}$ Data sharing is not applicable to this article as no new data were created or analyzed in this study.
	%\begin{definition}
	%	For a Hamiltonian vector field on a COLCS manifold $(M,\Omega,\theta,\eta)$,  vector field $Y$ is called a Noether symmetry if 
		%$$L_Y\Omega=di_Y\Omega+i_Y(\theta\wedge\Omega)=di_Y\Omega+\theta(Y)\Omega-\theta\wedge i_Y\Omega=d^\theta(i_Y\Omega)+\theta(Y)\Omega$$
		%$$i_{X_H}L_Y\Omega=-Y(H)+H\theta(Y)+\theta $$
	%\end{definition}
	
	%\begin{theorem}
	%	For a compact CoLCS manifold $(M,\Omega,\theta,\eta)$, there exists a non-degenerate pairing:
	%	\[
		%H^k_\theta(M) \times H^{2n+1-k}_{-\theta}(M) \to \mathbb{R}
		%\]
		%given by integration of forms.
	%\end{theorem}
    
%	\bibliographystyle{elsarticle-num}
	%\bibliography{references}
	
\end{document}